\documentclass[12pt,a4paper]{article}%

\usepackage{a4wide}
\usepackage{latexsym}
\usepackage{epsf}
\usepackage{amssymb}
\usepackage{graphicx}
\usepackage{amsmath, cite}
\usepackage{amsmath,amssymb,amsthm}
\usepackage{verbatim}
\usepackage{hyperref}%
\usepackage{amsmath}%
\usepackage{amsfonts}
\usepackage{xfrac}

\usepackage{tikz}

\renewcommand{\d}{\textrm{d}}

\newcommand{\w}{\wedge}

\newcommand{\sgn}{\textrm{sgn}\,}

\newcommand\varpm{\mathbin{\vcenter{\hbox{%
  \oalign{\hfil$\scriptstyle+$\hfil\cr
          \noalign{\kern-.3ex}
          $\scriptscriptstyle({-})$\cr}%
}}}}

\newcommand\varmp{\mathbin{\vcenter{\hbox{%
   \oalign{\hfil$\scriptstyle-$\hfil\cr
           \noalign{\kern-.3ex}
          $\scriptscriptstyle({+})$\cr}%
}}}}

\begin{document}
\numberwithin{equation}{section}
\begin{flushright}
\small

UUITP-15/13

\normalsize
\end{flushright}

\begin{center}
{\LARGE \textbf{{A note on M$2$-branes in opposite charge}}}

\vspace{1 cm} {\large Johan Bl{\aa}b{\"a}ck}\\

\vspace{0.85 cm}{Institutionen f{\"o}r fysik och astronomi,\\
Uppsala Universitet,\\  Box 803, SE-751 08 Uppsala, Sweden}
\\\vspace{0.4 cm}

\vspace{0.7cm} {\small\upshape\ttfamily johan.blaback @ physics.uu.se} \\

\vspace{4cm}


\textbf{Abstract}
\end{center}

\begin{quotation}
The flux singularity that arise when branes are put in an oppositely charged background is a wildly discussed issue and in this paper the problem is investigated further. New AdS world volume (anti-)M$2$-brane solutions are constructed by placing the brane in a background where the charge dissolved in the flux is opposite that of the brane. A topological obstruction is derived to show how the singularity develops during localisation of this solution and any similarly constructed solution. Furthermore the singularity is shown not to be possible to hide behind a horizon, which would have given credence to a possible M-theory resolution.
\end{quotation}

\newpage


\section{Introduction}

The interest of placing branes in backgrounds of oppositely charged flux comes from the now very famous idea of uplifting the energy of Anti-de Sitter space-times to de Sitter using anti-branes \cite{Kachru:2003aw}. The idea is to construct a stable AdS solution and then by placing a small amount of warped down anti-D3-branes one can achieve several phenomenologically interesting features -- supersymmetry (susy) breaking, and meta-stable arbitrarily small cosmological constant via the means of varying warping and the amount of branes.

These obviously desired phenomenological effects, from ten dimensional supergravity (sugra), the low energy theory of String Theory, is of great importance and have been studied extensively, for example the possibility of other decay channels \cite{Kachru:2002gs}. Not too long ago however, in an attempt to describe the backreaction of these anti-branes it was discovered that some modes developed singularities \cite{McGuirk:2009xx}. These singularities were something new because they were not directly sourced by the brane and appears in the flux surrounding the brane, not the field strength sourced by the brane.

This has spawned several discussions trying to explain and further investigate these flux singularities. Is the singularity really there in the supergravity solution or is it simply an artefact of partial smearing or treating backreaction perturbatively \cite{Bena:2009xk,Bena:2011hz,Bena:2011wh,Giecold:2011gw,Blaback:2011pn,Gautason:2013zw,Blaback:2011nz}? If the singularity is there, what does it mean or how can it be resolved \cite{Bena:2012tx,Bena:2012bk,Bena:2012vz,Blaback:2012nf,Bena:2013hr,Bena:2012ek}? So far neither partial smearing nor perturbative backreaction seems to be the issue that creates these singularities since the singularities develop even in fully localised solutions and in fully backreacted solutions \cite{Blaback:2011pn}, more general result in \cite{Gautason:2013zw}. To resolve the singularity a few methods have been tested. One such method is resolving the solutions via the Myers effect \cite{Myers:1999ps} \`{a} la Polchinski-Strassler \cite{Polchinski:2000uf}, by letting a D$(p+2)$-brane surround and cut off the singularity, created by a D$p$-brane, to a finite value. This method has so far been unsuccessful in resolving the singularity \cite{Bena:2012tx,Bena:2012bk,Bena:2012vz}. Another possibility that has been investigated is whether the singularity could signal for some new instability of the system, which still remains a possibility \cite{Blaback:2012nf}. The method of hiding the singularity behind an event horizon, as been successful for several systems\footnote{See \cite{Bena:2013hr} for an extensive list.}, would signal that the singularity could be resolved in String Theory, i.e. beyond the supergravity approximation. However when studied it has been shown and argued that hiding these type of singularities behind a horizon is not a possible means of getting rid of the singularity \cite{Bena:2013hr,Bena:2012ek}.

In the type II sugra it was also possible to give a possible interpretation as to the physics that gives rise to the singularity. As noticed in \cite{Blaback:2011pn}, elaborated upon in \cite{Blaback:2012nf}, the singularity arise in the flux charge density term ($H \w F_{6-p}$) of the Bianchi identity associated to a D$p$-brane, $\d F_{8-p} = H \w F_{6-p} + \textrm{sources}$. The sign of the flux charge density, integrated for compact or as a UV condition for non-compact internal spaces, should be opposite that of the brane to achieve the uplifting and susy-breaking properties sought. The oppositeness of the charge would make the flux attracted towards the brane a start clumping the flux to shield the charge. This lead to the interpretation of the singularity as arising because the Ans\"atze used were stationary while the physical system dynamical. In \cite{Blaback:2012nf} it was also discovered that the singularity could imply a new channel of instability for the KKLT vacua by possibly removing the barrier for meta-stability.

The same type of singularity has also been identified in $11$D supergravity \cite{Cottrell:2013asa,Giecold:2013pza,Massai:2011vi,Bena:2010gs}, but is not yet as comprehensively studied. For example there exists fully localised solutions, to linear order in a perturbative backreaction \cite{Cottrell:2013asa,Giecold:2013pza}, but a fully backreacted solution does not yet exist. The issue of trying to resolve the singularity in $11$D has also not yet completely kicked off yet -- but perhaps $11$D supergravity has the tools necessary to interpret the singularity.

In type II sugra several tools have, as mentioned, been invented to try and study these singularities. 
The purpose of this paper is to bring some of those tools into $11$D supergravity to investigate whether they provide some new information. One of these tools is what has been called a {\it topological no-go} \cite{Blaback:2011pn,Blaback:2011nz} which constrain the possible configurations of the potential of the field strength $F_{8-p}$ associated with a D$p$-brane and force a singularity in the flux density $H_3 \w F_{6-p}$.
The topological no-go have helped in several ways before. It helped to determine the presence of the singularity in a fully backreacted system of this type \cite{Blaback:2011pn,Blaback:2011nz}. The topological no-go also made it possible to give an interpretation of the flux singularity as flux polarisation \cite{Blaback:2011pn,Blaback:2012nf}, as explained previously.
 A similar topological constraint will be invented here, and from this it will be argued that it forces an unwanted singularity to develop in the flux. The other tool that will be used is putting in an event horizon and try to shield off the singularity. This has been argued not to be possible for the type II sugra singularities \cite{Bena:2013hr,Bena:2012ek} but perhaps M-theory possess the power to resolve these singularities while String Theory does not.


The study of backreaction in type II sugra has benefited from having available several more or less simple solutions. One such example in type II sugra is the anti-D$6$-brane solutions which helped the study of full backreaction \cite{Blaback:2011pn,Blaback:2011nz} and polarisation \cite{Bena:2012tx} in systems that display this flux singularity. Therefore this paper will also describe the construction of solutions similar to those previously used in type II sugra. These solutions are smeared space-filling (anti-)M$2$-brane solutions with AdS$_{3}$ world volume on a compact internal space.

The paper is organised as follows. Section \ref{sec:sugra} will present the conventions used for $11$D supergravity, and also go through the simplest form of the Ansatz used in the paper. Then Section \ref{sec:sols} will describe the new (anti-)M$2$-brane solution. This solution is constructed from space-filling (anti-)M$2$-branes with AdS$_3$ world volume on a compact internal manifold. Furthermore the solution utilises the approximation of smeared branes, hence the goal would be to localise the source. Because the branes are positioned into oppositely charged flux this is expected to become problematic, as experience would suggest. The purpose of Section \ref{sec:topo} is to derive a topological no-go that restricts the profile for the flux occupying the internal space. This no-go will force a singularity in the flux at the position of the brane. It will also be argued that this no-go is not only relevant for these particular solutions but should also be important to any study of branes located in oppositely charged flux. The no-go is extended in Subsection \ref{ssec:noblack} where a blackening-factor representing a horizon is introduced to shield the solution from this singularity. The same section also explains how this is not possible. Finally Section \ref{sec:con} summarises and concludes the paper.


\section{$11$D Supergravity}
\label{sec:sugra}

The conventions used are as follows. The $4$-form flux equation of motion and Bianchi identity
\begin{equation}
\begin{split}
\d \star_{11} G_4 &= \frac{1}{2} G_4 \w G_4 + Q \delta_8(\textrm{M}2),\\
\d G_4 &= 0.
\end{split}
\end{equation}
Here $Q = \varpm |Q|$ for the charge of the (anti-)M$2$-brane and the delta function is an $8$-form, $\delta_8(\textrm{M}2) = \delta(\textrm{M}2) \star_8 1$. The Einstein equations are
\begin{equation}
\begin{split}
R_{ab} &= \frac{1}{2}\left( |G_4|^2_{ab} - \frac{1}{3} |G_4|^2 g_{ab} \right) + \frac{1}{2}\left( T_{ab}^l - \frac{1}{9} T^l g_{ab} \right),\\
T_{ab}^l &= - T g_{\mu\nu} \delta(\textrm{M}2) \delta_{ab}^{\mu\nu},
\end{split}
\end{equation}
with $T = |T|$ being the tension of the brane and anti-brane. 

The Ansatz for the metric and the $4$-form flux is
\begin{equation}
\begin{split}\label{eq:metric}
\d s^2_{11} &= e^{2A} \d \tilde{s}^2_{2,1} + e^{2B} \d \tilde{s}^2_{8},\\
G_4 &= \star_{11} F_7 + F_4 + H_4,\\
F_7 &= e^{X} \star_{8} \d \alpha,\\
H_4 &= \lambda \star_8 F_4.
\end{split}
\end{equation}
The flux denoted $F_7$ is the field-strength that the brane sources, and the two four-form fluxes are the flux surrounding the brane. A parameter $\lambda$ is also introduced here to be able to vary the magnitude of the flux present in $H_4$ and $F_4$ which will turn out to be important later. This Ansatz is designed to have as much resemblance as possible to the type II sugra setups\cite{Blaback:2011pn,Blaback:2011nz,Blaback:2010sj}.
\begin{equation}
\begin{array}{|c|c|}
\hline
11 \textrm{D field} & \textrm{Type II analogue}\\
\hline
\hline
F_7 & F_{8-p}\\
F_4 & F_{6-p}\\
H_4 & H_3\\
\hline
\end{array}
\end{equation}

From this Ansatz to get the ``Fractional M$2$-brane'' solution \cite{Herzog:2000rz} simply put
\begin{equation}
\lambda = \varpm1,\ Q = \varpm|Q|,\ X = -3A,\ \alpha = \lambda e^{3A},\ B = -\frac{1}{2} A,
\end{equation}
where $\lambda = \varpm1$ implies that the combination $H_4 + F_4$ is (anti-)self dual ((A)SD), for (anti-)M$2$-branes.\\


\section{New non-SUSY AdS$_3$ solutions}
\label{sec:sols}

Presented here are new smeared solutions for M$2$-branes. These solutions are constructed in the same way as the anti-D$p$-branes in \cite{Blaback:2010sj}, and share some similarities in its construction.

The smeared approximation means that the M$2$-brane is spread out in the internal manifold. For the delta function this means that
\begin{equation}
Q \delta_8(\textrm{M}2) = Q \delta(\textrm{M}2) \star_8 1 = Q \tilde{\delta}(\textrm{M}2) \tilde{\star}_8 1\ \longrightarrow\ \frac{Q}{V} \tilde{\star}_8 1,
\end{equation}
where $V$ is the unwarped volume and tilde refers to the metric without warping or conformal factor (\ref{eq:metric}). From now on $V=1$, such that it is absorbed into the charge and tension. Furthermore, from the Ansatz used in the previous section, warping, conformal factor and the field $\alpha$ will also be removed
\begin{equation}\label{eq:off}
A \to 0,\ B\to 0,\ \alpha \to 0.
\end{equation}
The metric's internal part will then be split in two four dimensional parts, 
\begin{equation}
\d s^2_{11} = \d s^2_{2,1} + g^{(H)}_{ij}\d y^i \d y^j + g^{(F)}_{ij} \d z^i \d z^j,
\end{equation}
where $\d s^2_{2,1}$ is the world volume part of the metric, and ${}^{(H/F)}$ denotes that one is occupied by $H_4$ and one with $F_4$. 

Using the smearing Ansatz described above the equation of motion and Einstein equations now reduce to algebraic equations. The equation of motion gives a relation between the charge and the flux
\begin{equation}
0 = \lambda|F_4|^2 + Q.
\end{equation}
This is the tadpole cancellation condition and to solve this the signs of the charge and the flux need to be opposite $Q=\varpm |Q|$ and $\lambda = \varmp |\lambda|$, which would correspond to (anti-)M$2$-branes in oppositely charged flux.

\begin{figure}[h!]
\begin{center}
\includegraphics{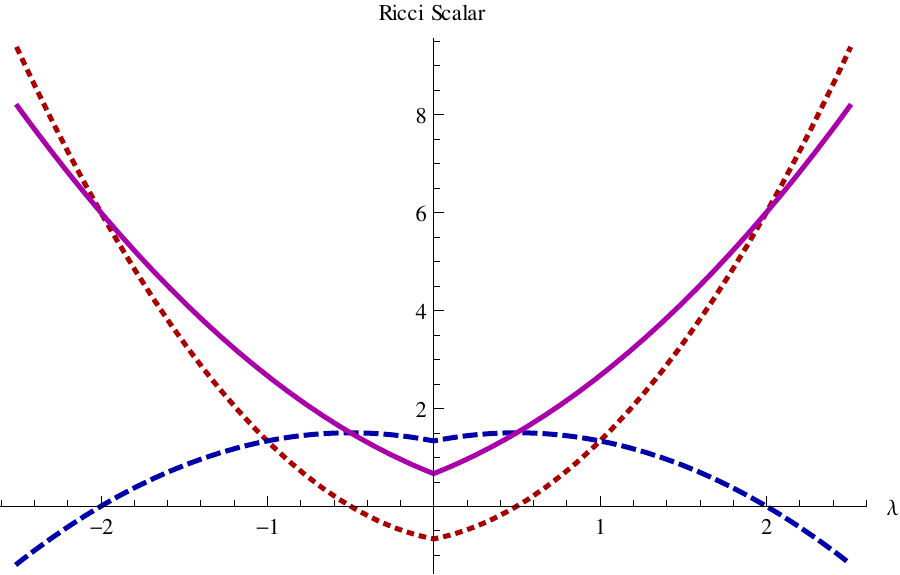}
\end{center}
\caption{The curvature of the ${}^{(H)}$ section is the dashed red line and the ${}^{(F)}$ section is the dotted blue line. The full purple line is the total curvature of the internal space. The left (right) side of the vertical axis is where $Q$ is positive (negative) and $\lambda$ is negative (positive).\label{fig:curv}}
\end{figure}

The Einstein equations will now give the following expressions. Externally the curvature is necessarily negative
\begin{equation}
R_{\mu\nu} = -\frac{1}{3}\left( \frac{1+\lambda^2}{2} |F_4|^2 + T_{\textrm{M}2}\right) g_{\mu\nu}.
\end{equation}
The curvature of the internal spaces can vary and even switch sign
\begin{equation}
\begin{split}
R^{(H)}_{ij} &= \frac{1}{6}\left( (2\lambda^2 - 1)|F_4|^2 + T_{M2}\right) g^{(H)}_{ij} = \frac{1}{6} (-1+|\lambda| +2 \lambda ^2)|F_4|^2 g^{(H)}_{ij}, \\
R^{(F)}_{ij} &= \frac{1}{6}\left( (2 - \lambda^2)|F_4|^2 + T_{M2}\right) g^{(F)}_{ij} = \frac{1}{6} (2+|\lambda| - \lambda ^2)|F_4|^2 g^{(F)}_{ij},
\end{split}
\end{equation}
having used the tadpole condition and that the tension is necessarily positive for (anti-)M$2$-branes. Notice here that $\lambda$ is a continuous parameter. Choosing different values for $\lambda$ the curvatures varies according to Figure \ref{fig:curv}, where it should also be noted that the total curvature of the internal space remains positive for all values of $\lambda$. In the case of AdS$_{p+1}$ solutions for the D$p$ branes of \cite{Blaback:2010sj} this is not the case, since there the dilaton equation of motion restricts $\lambda$ to a fixed value.\footnote{In \cite{Blaback:2010sj} $\kappa = 1/\lambda$.} It should also be noted that when a circle can be identified in the ${}^{(H)}$ section of the internal space this construction corresponds to a direct uplift from type IIA sugra to $11$D sugra. One would expect these to be stable to some degree since the analogue solution for anti-D$6$-branes with AdS$_7$ world volume were discovered to be stable with respect to the left-invariant closed string moduli \cite{Blaback:2011nz}.

The above construction is, as mentioned, fully smeared. To continue with localisation one have to turn warping, conformal factor and $\alpha$ back on. The profile of the flux distribution in the internal manifold should also vary. For the fractional brane solutions \cite{Herzog:2000rz} the (anti)-M$2$-branes are positioned in (A)SD which is the ``natural background'', or \emph{BPS} in the same sense as declared in \cite{Blaback:2010sj}, for the brane. When the brane is forced into a compact background as considered here, where the charge dissolved in fluxes around the brane and the brane itself are not mutually BPS, the flux will arrange it self into new configurations. This means that $\lambda$ will be promoted to a function that describes the flux distribution in the localisation direction.


\section{The topological no-go}
\label{sec:topo}

Consider the similar division of the internal manifold as earlier
\begin{equation}
M_8 = (\mathbb{R} \times M_3) \times M_4,
\end{equation}
where, equivalent to the type II sugra constructions, $H_4$ occupies the $\mathbb{R} \times M_3$ section, and $F_4$ occupies $M_4$. Here $z$ is introduced to parametrise $\mathbb{R}$ and localisation will only be considered along this direction and potentially still remain smeared along other directions. The metric has of the form
\begin{equation}
\d s^2_{11} = e^{2A}\d s^2_{2,1} + e^{2B}(\d z^2 + h(z)^2 \d \tilde{s}_3^2 + \d \tilde{s}_4^2),
\end{equation}
where the factor $h(z)^2$ is present to show that this will work for compact (e.g. $h(z)=\sin(z)$), such as the smeared solution presented previously, and non-compact (e.g. $h(z)=z$) spaces.

Considering the exclusively internal legs of the equation of motion for $G_4$ one acquires the following expression
\begin{equation}\label{eq:bianchi}
e^{-3A+6B}\partial_z^2 \alpha + \partial_z \alpha \partial_z (e^{-3A+6B}h^{3}) h^{-3}= \lambda |\tilde{F}_4|^2 + Q \tilde{\delta}(M2) h^{-3}.
\end{equation}
There is also a relation that relates $\alpha$ and $\lambda$, this is derived from the equations of motion for $G_4$ considering the external legs
\begin{equation}\label{eq:alrel}
\alpha = \lambda e^{3A}.
\end{equation}
Note that this forces $\alpha$ and $\lambda$ to have the same sign and the fact that $e^{3A}$ tends to zero at the origin -- which is usually considered for a brane -- hence a finite $\alpha$ at the origin implies a singularity in $\lambda$ of same sign. This is an important point that we will get back to. Considering (\ref{eq:bianchi}) at an extremal point for the function $\alpha$ and substituting $\lambda$ for $\alpha$ using (\ref{eq:alrel}), the resulting expression is
\begin{equation}\label{eq:nogo}
\left. \sgn \alpha = \sgn \alpha''\quad \right|_{\alpha'=0}.
\end{equation}
These two conditions will make it sufficient to see that there is a singularity in $\lambda$.

The problem at hand is to place (anti-)M$2$-branes, $\sgn Q = \varpm 1$, in a background that is not BPS in relation to the (anti-)brane, i.e. $\sgn \lambda = \varmp 1$. Even though the branes are placed in oppositely charged flux, the flux do not need to have the same charge through out the whole internal space. In fact the only thing that will be imposed is $\sgn \lambda = \varmp 1$ as a UV condition, i.e. to be true at $z\to \infty$ (or corresponding point on a compact space). The UV condition is important both compact and non-compact. Consider the smeared solution presented in the previous section, which is on a compact internal space. Smearing can be interpreted as using the integrated equations, which for the source terms means that the delta function is replaced by a constant and similarly for the flux, i.e. that the function $\lambda(z)$ is replaced by its integrated constant value. If the smeared solution is suppose to correspond to the localised one, which is commonly the assumption one is working under using the smeared approximation, this integral have to be equal to that constant. This is what the UV condition assures. For non-compact spaces this UV condition is the same expected from uplifting procedures as in KKLT, that is a brane is placed in a background that is not mutually BPS and the boundary conditions are assumed to be the same to be able to ``glue'' the non-compact space onto a compact one.

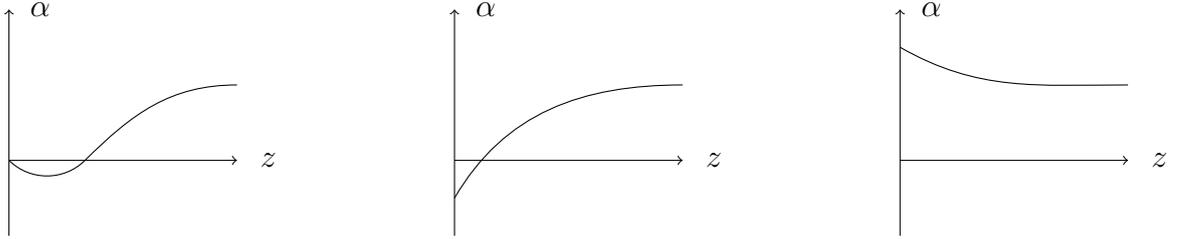
\begin{figure}
\begin{center}
\begin{tabular}{ccc}
\begin{tikzpicture}[scale=.5]
\coordinate (origin) at (0,2);
\coordinate (yend) at (0,6);
\coordinate (ystart) at (0,0);
\coordinate (xend) at (6,2);
\draw [->] (ystart) -- (yend);
\draw [->] (origin) -- (xend);
\node (yendn) at (yend) {$\qquad \alpha$};
\node (xendn) at (xend) {$\qquad z$};
\coordinate (infinity) at (6,4);
\coordinate (midpoint) at (2,2);
\draw (origin) to[out=-45,in=-135,thick] (midpoint);
\draw (midpoint) to[out=45,in=180,thick] (infinity);
\end{tikzpicture}
$\quad$
&
$\quad$
\begin{tikzpicture}[scale=.5]
\coordinate (origin) at (0,2);
\coordinate (yend) at (0,6);
\coordinate (ystart) at (0,0);
\coordinate (xend) at (6,2);
\draw [->] (ystart) -- (yend);
\draw [->] (origin) -- (xend);
\node (yendn) at (yend) {$\qquad \alpha$};
\node (xendn) at (xend) {$\qquad z$};
\coordinate (infinity) at (6,4);
\draw (0,1) to[out=60,in=180,thick] (infinity);
\end{tikzpicture}
$\quad$
&
$\quad$
\begin{tikzpicture}[scale=.5]
\coordinate (origin) at (0,2);
\coordinate (yend) at (0,6);
\coordinate (ystart) at (0,0);
\coordinate (xend) at (6,2);
\draw [->] (ystart) -- (yend);
\draw [->] (origin) -- (xend);
\node (yendn) at (yend) {$\qquad \alpha$};
\node (xendn) at (xend) {$\qquad z$};
\coordinate (infinity) at (6,4);
\coordinate (start) at (0,5);
\draw (start) to[out=-30,in=180,thick] (infinity);
\end{tikzpicture}
\end{tabular}
\end{center}
\caption{The left sketch is excluded by (\ref{eq:nogo}), the middle is excluded by (\ref{eq:bc}), the right remains allowed and implies flux that is singular an not mutually BPS with the brane.\label{fig:3pics}}
\end{figure}

The first construction that one would like to consider is to have the flux being mutually BPS with the brane at the position of the brane, $z=0$. This will give the following conditions
\begin{equation}
\textrm{{\bf 1.}}\qquad \textrm{IR: } \alpha \to \varpm 0,\ \quad\textrm{UV: } \alpha \to \varmp |\alpha_\infty|,
\end{equation}
where $\alpha_\infty$ is simply the value that $\alpha$ would limit to at infinity. The sketch of such a profile is the leftmost picture in Figure \ref{fig:3pics} (only drawn for anti-M$2$-branes). By looking at the extremal points one gets from connecting the IR and UV boundary conditions are exactly of the type that are excluded by (\ref{eq:nogo}).

To change the boundary conditions to evade restriction (\ref{eq:nogo}) the IR boundary conditions have to be modified. This can be done in the following
\begin{equation}
\textrm{{\bf 2.}}\qquad \textrm{IR: } \alpha \to \varpm |\alpha_0|\ \ \&\ \ \alpha' \to \varmp |\alpha_0'|,\ \quad\textrm{UV: } \alpha \to \varmp |\alpha_\infty|.
\end{equation}
These boundary conditions would imply that the (anti-)brane is surrounded by a singular flux of same sign charge. The corresponding sketch would be the middle picture of Figure \ref{fig:3pics} (again only drawn for anti-M$2$-branes). This sketch is however not allowed. If integrated, (\ref{eq:bianchi}), where the left hand side is a total derivative, gives the following expression
\begin{equation}\label{eq:bc}
\left.e^{-3A + 6B} h^3 \alpha' \right|_{z\textrm{-const.}} = Q,
\end{equation}
which is valid for the $z$-constant term in an expansion, and hence dominating at $z=0$, assuming $\alpha$ is finite there. This means that the only allowed IR conditions here is $\sgn \alpha' = \sgn Q$ which was not obeyed.

This leaves only one remaining option
\begin{equation}
\textrm{{\bf 3.}}\qquad \textrm{IR: } \alpha \to \varpm |\alpha_0|\ \ \&\ \ \alpha' \to \varpm |\alpha_0'|,\ \quad\textrm{UV: } \alpha \to \varmp |\alpha_\infty|.
\end{equation}
The above boundary condition is the only one that evades all restrictions and is the only possible candidate. The sketch is found as the rightmost picture in Figure \ref{fig:3pics}, (anti-M$2$-branes). These boundary conditions do imply that there is a singular flux surrounding the brane, which has opposite charge.

\subsection{No blackening of M$2$ branes}
\label{ssec:noblack}

The metric can be extended to include a {\it blackening-factor}, $e^{2f(z)} = 1 - \sfrac{|k|}{z}$, that has the possibility to hide the forced singularity from above, behind a horizon. This would provide a possibility to resolve the singularity in M-theory. The new metric is
\begin{equation}
\d s^2_{11} = e^{2A} \left( - e^{2f} \d t^2 + \d x^2_2 \right) + e^{2B} \left( e^{-2f} \d z^2 + g(z)^2 \d \tilde{s}^2_3 + e^{2C} \d \tilde{s}^2_4\right),
\end{equation}
the function $C=C(z)$ has also been added for generality. This metric gives a similar expression as (\ref{eq:bianchi}) from the $G_4$ equation of motion
\begin{equation}
e^{-3A+6B+f} \partial_z^2 \alpha + \partial_z \alpha \partial_z \left( e^{-3A+6B + 4C} h^3 \right) e^{f-4C} h^{-3} = \lambda e^{-8C} |\tilde{F}_4|^2 + Q_{M2} \tilde{\delta}(M2)e^{f-4C}h^{-3}.
\end{equation}
This expression still provides the starting point for the above no-go
\begin{equation}
\left.\sgn \alpha'' = \sgn \alpha\right|_{\textrm{extrema}},\quad \alpha = \lambda e^{3A+f} , \quad e^{-3A + 6B +4C} h^3 \partial_z \alpha|_{z-\textrm{const.}} = Q.
\end{equation}
The flux Ansatz is still the same, with $X= -3A -f$. It is still so that one can only have $\sgn \alpha'' = \sgn \alpha$ extremal points, the blackening-factor has now taken the the role that the warp factor $A$ had before, since now $e^{2f}$ tends to zero at the horizon and gives a singularity to $\lambda$, now simply pushed in front of the horizon.


\section{Conclusions}
\label{sec:con}

In this paper new AdS$_3$ solutions were presented. These solutions are (anti-)M$2$-brane solutions on a compact manifold and the sign of the total curvature of the internal manifold is positive, however depending on the flux Ansatz used one section of the internal manifold can have negative curvature. To solve the so called tadpole condition, the charge dissolved in the fluxes around the brane must be opposite that of the brane itself to create net zero charge internally. In type II sugra similar setups have been studied and present a singularity in the surrounding flux that is not directly sourced by the brane. Hence it should be expected that also the smeared (anti-)M$2$-brane solutions presented here will suffer from the same problem. This was also demonstrated in the paper. By localising the brane in one direction a topological restriction to the profile of the flux distribution is derived. It was furthermore shown how this restriction forces a singularity in the flux density to develop. The topological argument was also extended to include a blackening factor that creates a horizon around the brane, in an attempt to hide the singularity for a possible M-theory resolution of it. In exactly the same manner as a singularity is forced to develop in case of no horizon, the horizon simply shifts the singularity to appear at its surface.

The localisation problem that is described in this paper is mainly studied from the point of view of the solutions also presented within. However the argument as to why a singularity develops is more general and has in the type II sugra systems shown to give a good general picture and understanding for these types of singularities. This result do indicate that the singularity should be present even after full backreaction.


\section*{Acknowledgments}

The author would like to thank Thomas Van Riet for discussions, motivation and suggesting this project, Ulf Danielsson for comments and suggestions on draft and layout and Lena Heijkenskj\"old for comments on the draft. The author is supported by the G{\"o}ran Gustafsson Foundation.


\bibliography{refs}
\bibliographystyle{utphysmodb}

\end{document}